\documentclass[prd,aps,showpacs,nofootinbib,showkeywords,eqsecnum]{revtex4-1}

\usepackage{graphicx,color,amsmath,amsxtra}
\usepackage{epsf}
\usepackage{amssymb}
\usepackage{enumerate}
\usepackage{hhline}
\usepackage{array}
\usepackage{tabularx}
\usepackage[unicode]{hyperref}
\usepackage{graphicx}                
\usepackage{epstopdf}

\begin{document}
\pagestyle{myheadings}

\title{No-go theorem for inflation in an extended Ricci-inverse gravity model}
\author{Tuan Q. Do }
\email{tuan.doquoc@phenikaa-uni.edu.vn}
\affiliation{Phenikaa Institute for Advanced Study, Phenikaa University, Hanoi 12116, Vietnam}
\affiliation{Faculty of Basic Sciences, Phenikaa University, Hanoi 12116, Vietnam}
\date{\today} 

\begin{abstract}
In this paper, we propose an extension of the Ricci-inverse gravity, which has been proposed  recently as a very novel type of fourth-order gravity, by introducing a second order term of the so-called anticurvature scalar as a correction. The main purpose of this paper is that we would like to see whether the extended Ricci-inverse gravity model admits the homogeneous and isotropic Friedmann-Lemaitre-Robertson-Walker  metric as its stable inflationary solution. However, a no-go theorem for inflation in this extended Ricci-inverse gravity is shown to appear through a stability analysis based on the dynamical system method. As a result, this no-go theorem implies that it is impossible to have such stable inflation in this extended Ricci-inverse gravity model.
\end{abstract}

%


\maketitle
\section{Introduction} \label{intro}
Cosmic inflation \cite{Starobinsky:1980te,Guth:1980zm,Linde:1981mu} has been regarded as one of the leading paradigms for modern cosmology. Remarkably, many theoretical predictions derived within the context of  cosmic inflation have been shown to be highly consistent with the observed data of the leading cosmic microwave background radiation (CMB) detectors such as the Wilkinson Microwave Anisotropy Probe satellite (WMAP) ~\cite{WMAP} and the  Planck one~\cite{Planck}.

It is widely believed that the inflationary phase of our universe, which happens very shortly after the Big Bang, is driven by the so-called inflaton field, which is a hypothesis scalar field \cite{Guth:1980zm,Linde:1981mu}. The nature of this scalar field, however, has been a great mystery of modern cosmology. It appears that many inflation models have been proposed to realize the origin of the inflaton field, e.g., see Ref. \cite{string}. In addition, many inflation models have been examined  their cosmological viability through comparing with the observational data of the Planck satellite, e.g., see Ref. \cite{Martin:2013tda}. It is worth noting that among the well-known inflation models the Starobinsky model  \cite{Starobinsky:1980te}, one of the first inflation models, involving the $R^2$ correction term, has remained as one of the most favorable models in the light of the Planck observation \cite{Planck}. Although the Starobinsky model originally contains no scalar field, but it has been shown that the Starobinsky model can be conformally transformed into an effective model of scalar field \cite{Whitt:1984pd,Barrow:1988xh,Maeda:1987xf,Muller:1989rp,Mishra:2019ymr}. More interestingly, according to Ref. \cite{Salvio:2018crh}  the Starobinsky model can become, under a suitable Weyl transformation, an effective gravity model involving a more general matter sector, which contains not only scalar field but also other fields.  It is noted that the Starobinsky model is one of the simplest fourth-order  gravity models (a.k.a. quadratic gravity), where the cosmic inflation can be found \cite{Barrow:1983rx,Starobinsky:1987zz,Mijic:1987bq,barrow05,barrow06,Middleton:2010bv,kao09,Toporensky:2006kc,Muller:2017nxg,Koshelev:2017tvv}. The other types of the fourth-order gravity such as $R_{\mu\nu}R^{\mu\nu}$  and  $R_{\mu\nu\alpha\beta}R^{\mu\nu\alpha\beta}$ can be seen in Refs. \cite{barrow05,barrow06,Middleton:2010bv}. It is also noted that the fourth-order gravity has been one of the leading alternative approaches to solve the so-called accelerated expansion problem of the current universe \cite{DeFelice:2010aj,Nojiri:2010wj,Nojiri:2017ncd,Carroll:2004de,Amendola:2006kh,Myrzakulov:2015qaa}. More interestingly, inspired by the fourth-order gravity a number of interesting gravity models have been proposed to unify both the early and late time phases of our universe into a single scenario, see, e.g., Refs. \cite{Nojiri:2010wj,Myrzakulov:2015qaa}. All these facts indicate that the fourth-order gravity has been one of the most attractive frameworks for studying both the early and late time phases of our universe. For an interesting review on the rich history and cosmological implications of the fourth-order gravity, see Ref. \cite{Schmidt:2006jt}. In addition to the cosmological aspects, it is worth noting that gravitational actions including terms quadratic in the curvature tensor such as $R^2$ and $R_{\mu\nu}R^{\mu\nu}$ have been shown to be renormalizable by Stelle \cite{Stelle:1976gc}. This important result indicates that the fourth-order gravity is a promising approach to quantum gravity despite the fact that it could admit the Ostrogradsky ghost \cite{Woodard:2015zca} due to the existence of higher derivatives. It is noted that the Starobinsky model turns out to be free of the Ostrogradsky ghost \cite{Woodard:2015zca}. For detailed discussions on the interesting issues related to the quantum scenario of the fourth-order gravity such as the renormalizability,  ghost problem, Landau poles, etc., see an interesting review \cite{Salvio:2018crh}.

Recently, a very novel fourth-order gravity model, which is called the Ricci-inverse gravity, has been proposed by Amendola, Giani, and Laverda  in Ref. \cite{Amendola:2020qho}.  This model is constructed by introducing a very novel geometrical object called an {\it anticurvature scalar} denoted by the capital letter $A$.  More specific, the anticurvature scalar $A$ is nothing but the trace of {\it anticurvature tensor} denoted as $A^{\mu\nu}$, which is defined to be equal to the inverse Ricci tensor, i.e.,  $A^{\mu\nu} =R_{\mu\nu}^{-1}$. Very interestingly,  this model has been shown to admit a no-go theorem stating that both decelerated and accelerated expansions cannot exist together in this model. Consequently, it is impossible for the Ricci-inverse gravity model to be a dark energy candidate \cite{Amendola:2020qho}. One therefore might think of a possibility that the Ricci-inverse gravity may be suitable for describing the early time inflationary phase rather than the late time accelerated expansion phase of the universe \cite{Amendola:2020qho}. However, our follow-up study in Ref. \cite{Do:2020vdc} has indicated that we also have another no-go theorem, not for accelerated expansion but for inflation, of the Ricci-inverse gravity model. In particular, we have shown that although the original Ricci-inverse gravity model admits both the Friedmann-Lemaitre-Robertson-Walker (FLRW) and Bianchi type I solutions but all of these solutions turn out to be unstable against perturbations during the inflationary phase. All of these results have raised doubts about the cosmological viability of the Ricci-inverse gravity. Hence, non-trivial extensions such as the $f(R,A)$ seem to be necessary to cure the Ricci-inverse gravity \cite{Amendola:2020qho}. 

In fact, some simple extensions of the Ricci-inverse gravity have  been proposed in Ref.  \cite{Amendola:2020qho} in order to overcome the first no-go theorem. Some of them turn out to be promising and of course need to be verified by further investigations. In this paper, motivated by the Starobinsky model \cite{Starobinsky:1980te} we would like to propose a simple extension of the Ricci-inverse gravity by introducing a second order term $A^2$. Note that the introduction of $A^2$ is not for violating the first no-go theorem for the accelerated expansion. Therefore, it has not been proposed in Ref. \cite{Amendola:2020qho}. By doing this, we  expect to have the corresponding isotropic inflation solution, which would be stable against perturbations. Unfortunately, a no-go theorem based on the stability analysis will be shown to hold for the isotropic inflation. This result together with our previous investigation \cite{Do:2020vdc} raise more doubt about the implications of the Ricci-inverse gravity for the inflationary phase of our universe. 

As a result, this paper will be organized as follows: (i) A brief introduction of the present study has been written in the Sec. \ref{intro}. (ii) Basic setup of the extended Ricci-inverse gravity model will be presented in Sec. \ref{sec2}.  (iii) Isotropic inflationary solution to this model will be solved analytically in Sec. \ref{sec3}. (iv) The proof of the no-go theorem for this inflationary solution will be shown in Sec. \ref{sec4}. (v) Finally, concluding remarks will be written in Sec. \ref{final}.
\section{Basic setup} \label{sec2}
As a result, an action of extended Ricci-inverse gravity has been proposed in Ref. \cite{Amendola:2020qho} as follows
\begin{equation} \label{action}
S=\int d^4 x \sqrt{-g}  \left[R+ f(A) -2\Lambda\right],
\end{equation}
where the reduced Planck mass, $M_p$, has been set to be one for convenience, while $\Lambda >0 $ is the pure cosmological constant \cite{barrow06,Toporensky:2006kc,Muller:2017nxg}. In addition, $f(A)$ is an arbitrary  function of the so-called anticurvature scalar, $A$, which is nothing but the trace of the so-called anticurvature tensor, $A^{\mu\nu}$. By construction, the anticurvature tensor is nothing but the Ricci-inverse tensor \cite{Amendola:2020qho}, i.e., 
\begin{equation}
A^{\mu\nu} =R_{\mu\nu}^{-1}.
\end{equation}
It is noted that this relation does not lead to $A=R^{-1}$.  As a result, varying the action \eqref{action} with respect to $g^{\mu\nu}$ will yield the corresponding Einstein field equation \cite{Amendola:2020qho},
\begin{equation} \label{Einstein-field-equation}
R^{\mu\nu}-\frac{1}{2}Rg^{\mu\nu} +\Lambda g^{\mu\nu} =  f_A  A^{\mu\nu} + \frac{1}{2} f g^{\mu\nu} - \frac{1}{2} \left[ 2g^{\rho\mu}\nabla_\alpha \nabla_\rho f_A A^{\alpha}_\sigma A^{\nu\sigma} -\nabla^2 \left( f_A A^{\mu}_\sigma A^{\nu\sigma}\right) -g^{\mu\nu}\nabla_\alpha \nabla_\rho \left(f_A A^\alpha_\sigma A^{\rho\sigma} \right)\right],
\end{equation}
where  $f_A =\partial f/\partial A$ and $\nabla_\mu$ is understood as the covariant derivative.

It turns out that the right hand side of Eq. \eqref{Einstein-field-equation} looks very complicated which addresses a very lengthy calculation, even for the simplest metrics such as the Friedmann-Lemaitre-Robertson-Walker (FLRW) metric \cite{Amendola:2020qho}. For convenience, therefore, we decide to use another approach based on the Euler-Lagrange equations, which will turn out to be very effective in computing explicitly the non-vanishing components of the Einstein field equation. Of course, this method has also been used in our previous study \cite{Do:2020vdc}. As a result, the calculation process is as follows: (i) given a specific metric we will first define the corresponding Lagrangian of an extended Ricci-inverse gravity model, i.e., 
\begin{equation}
{\cal L} = \sqrt{-g} \left[R+f(A) -2\Lambda \right],
\end{equation}
(ii) then we will define the corresponding Euler-Lagrange equations of scale factors, which are nothing but the desired field equations.

 In this paper, motivated by the Starobinsky gravity  \cite{Starobinsky:1980te}, we propose to study a simple generalization of the Ricci-inverse gravity, in which $f(A)$ contains not only $A$ but also the second order term $A^2$, i.e.,
 \begin{equation}
  f(A)=\alpha_1 A +\alpha_2 A^2,
  \end{equation}
  where $\alpha_1$ and $\alpha_2$ are free parameters. We expect that the inclusion of the higher order term $A^2$ would lead to  stability regions for the existing of isotropic inflationary solutions.

In this paper, we will consider the spatially homogeneous and isotropic FLRW spacetime described by the following  metric,
\begin{equation} \label{metric}
ds^2 =-N^2(t)dt^2 +\exp[2\beta(t)]\left(dx^2+dy^2 +dz^2 \right),
\end{equation}
where  $N(t)$ is the lapse function, whose existence is necessary to derive the corresponding Friedmann equation from its Euler-Lagrange equation \cite{Toporensky:2006kc,Kao:1991zz}. In particular, after deriving the corresponding Euler-Lagrange equation equation, $N$ will be set as one in order to recover the  well-known Friedmann equation \cite{Toporensky:2006kc,Kao:1991zz}. In addition, $\beta(t)$ is an isotropic scale factor, which is assumed to be a function of cosmic time $t$ due to the homogeneity of the FLRW spacetime. As a result, the corresponding non-vanishing components of the Ricci tensor, $R_{\mu\nu}\equiv R^\rho{ }_{\mu\rho\nu}$, can be defined to be
\begin{align}
R_{00}&= 3 \left( \frac{\dot N}{N} \dot\beta - \Phi \right), \\
R_{33}=R_{22}=R_{11}&= -\frac{g_{11}}{N^2} \left( \frac{\dot N}{N} \dot\beta - \Pi \right),\end{align}
respectively. Here, two additional variables $\Phi$ and $\Pi$ have been defined as
\begin{align}
\Phi &= \ddot\beta +\dot\beta^2,\\
\Pi &=\ddot\beta +3\dot\beta^2,
\end{align}
respectively, for convenience. Note that $\dot\beta \equiv d\beta/dt$, $\ddot\beta \equiv d^2\beta/dt^2$, $\beta^{(3)} \equiv d^3\beta/dt^3$, and so on.
Thanks to these useful results, the corresponding Ricci scalar,  $R \equiv g^{\mu\nu}R_{\mu\nu}$, and anticurvature scalar, $A\equiv g_{\mu\nu}A^{\mu\nu}$,  turn out to be
\begin{align} \label{def-of-R}
R =& -\frac{6}{N^2}\left[  \frac{\dot N}{N} \dot\beta -  \left(\ddot\beta +2\dot\beta^2 \right) \right],\\
\label{def-of-A-1}
A  =& - \frac{N^2}{3} \left( \frac{\dot N}{N} \dot\beta - \Phi \right) ^{-1} -3{N^2 } \left( \frac{\dot N}{N} \dot\beta - \Pi \right)^{-1},
\end{align}
respectively. This result confirms the above claim that $A\neq R^{-1}$. As a result, the Lagrangian,
\begin{equation}
 {\cal L}= \exp[3\beta] N\left(R+\alpha_1 A+\alpha_2 A^2-2\Lambda \right),
 \end{equation} 
 can now be defined explicitly. It appears that this Lagrangian explicitly contains two independent variables, $N(t)$ and $\beta(t)$, and their time derivatives. Next, we are going to figure out the corresponding Euler-Lagrange  equations in order to figure out cosmological solutions. First, the following Euler-Lagrange equation for the lapse function $N$ is defined as follows
 \begin{equation}
\frac{\partial {\cal L}}{\partial N} - \frac{d}{dt}\left( \frac{\partial {\cal L}}{\partial \dot N} \right)=0,
\end{equation}
which will become, after setting $N=1$, the corresponding Friedmann equation,\begin{align} \label{field-equation-1}
&\frac{\alpha_1}{3\Phi} \left[ 3 -\frac{1}{\Phi} \left(\ddot\beta +3\dot\beta^2 \right) +\frac{2}{\Phi^2} \dot\beta \left(\beta^{(3)} +2\dot\beta \ddot\beta \right) \right] +\frac{3\alpha_1}{\Pi} \left[2 + \frac{2}{\Pi^2}\dot\beta \left(\beta^{(3)}+6\dot\beta \ddot\beta \right) \right] \nonumber\\
&+\frac{\alpha_2 }{9\Phi^2}\left[ 5 -\frac{2}{\Phi} \left(\ddot\beta+3\dot\beta^2 \right) +\frac{6}{\Phi^2}\dot\beta \left(\beta^{(3)}  + 2\dot\beta \ddot\beta \right)  \right] +\frac{9\alpha_2}{\Pi^2} \left[ 3+\frac{6}{\Pi^2} \dot\beta \left(\beta^{(3)} +6\dot\beta\ddot\beta \right) \right] \nonumber\\
&+\frac{\alpha_2}{\Pi \Phi} \left[5 -\frac{2}{\Phi} \left(\ddot\beta +3\dot\beta^2 \right) +\frac{4}{\Phi^2} \dot\beta \left(\beta^{(3)} +2\dot\beta \ddot\beta \right) +\frac{2}{\Pi \Phi} \dot\beta \left(\beta^{(3)} +4\dot\beta \ddot\beta \right) \right] \nonumber\\
& +\frac{\alpha_2}{\Phi \Pi} \left[3 +\frac{4}{\Pi^2} \dot\beta \left(\beta^{(3)} +6\dot\beta \ddot\beta \right) +\frac{2}{\Phi \Pi} \dot\beta \left(\beta^{(3)} +4\dot\beta \ddot\beta \right) \right] \nonumber\\
&+ 2\left(3 \dot\beta^2 -\Lambda\right)=0.
\end{align}
On the other hand, the corresponding Euler-Lagrange equation of $\beta$ turns out to be
\begin{equation} \label{EL-2}
\frac{\partial {\cal L}}{\partial \beta} - \frac{d}{dt} \left(\frac{\partial {\cal L}}{\partial \dot \beta} \right)+  \frac{d^2}{dt^2}\left( \frac{\partial {\cal L}}{\partial \ddot \beta} \right)=0,
\end{equation}
which will be reduced, after setting  $N=1$, to
\begin{align} \label{field-equation-2}
&\frac{\alpha_1}{3\Phi} \left\{ 3 -\frac{1}{\Phi} \left(\ddot\beta +3\dot\beta^2 \right) +\frac{2}{\Phi^2}  \left[\beta^{(4)}+6\dot\beta \beta^{(3)} +2 \ddot\beta \left(\ddot\beta+4\dot\beta^2 \right) \right] - \frac{6}{\Phi^3} \left(\beta^{(3)} +2\dot\beta \ddot\beta \right)^2 \right\} \nonumber\\
&+\frac{3\alpha_1}{\Pi} \left[6+ \frac{2}{\Pi^2} \left(\beta^{(4)}+6\dot\beta \beta^{(3)}+6 \ddot\beta^2 \right) - \frac{6}{\Pi^3} \left(\beta^{(3)} +6\dot\beta \ddot\beta \right)^2 \right] \nonumber\\
&+\frac{\alpha_2 }{9\Phi^2}\left\{ 3 -\frac{2}{\Phi} \left( \ddot\beta+ 3\dot\beta^2 \right) +\frac{6}{\Phi^2} \left[ \beta^{(4)} +6\dot\beta \beta^{(3)} + 2\ddot\beta \left( \ddot\beta+ 4  \dot\beta^2 \right)\right] -\frac{24}{\Phi^3} \left(\beta^{(3)} +2\dot\beta\ddot\beta \right)^2  \right\} \nonumber\\
&+\frac{27\alpha_2}{\Pi^2} \left[ 3 +\frac{2}{\Pi^2} \left(\beta^{(4)}+6 \dot\beta \beta^{(3)} +6 \ddot\beta^2  \right)  -\frac{8 }{\Pi^3}  \left(\beta^{(3)} +6\dot\beta\ddot\beta \right)^2 \right] \nonumber\\
& +\frac{\alpha_2}{\Pi \Phi} \left\{ 3 -\frac{2}{\Phi} \left(\ddot\beta +3\dot\beta^2 \right) +\frac{4}{\Phi^2} \left[\beta^{(4)} +6\dot\beta \beta^{(3)} +2 \ddot\beta \left(\ddot\beta +4\dot\beta^2\right) \right] -\frac{12}{\Phi^3} \left(\beta^{(3)} +2\dot\beta \ddot\beta \right)^2 \right. \nonumber\\
&\left.+\frac{2}{\Pi \Phi}  \left[\beta^{(4)} +6\dot\beta \beta^{(3)}+4 \ddot\beta \left(\ddot\beta+3\dot\beta^2 \right) \right] -\frac{4}{\Pi \Phi^2} \left(\beta^{(3)} +2\dot\beta \ddot\beta \right)\left( 3\beta^{(3)} +14 \dot\beta \ddot\beta \right) \right\} \nonumber\\
&+\frac{\alpha_2}{\Phi \Pi} \left\{9 +\frac{4}{\Pi^2}  \left(\beta^{(4)} +6\dot\beta \beta^{(3)} +6\ddot\beta^2 \right) -\frac{12}{\Pi^3} \left(\beta^{(3)} +6\dot\beta \ddot\beta \right)^2 \right. \nonumber\\
&\left. +\frac{2}{\Phi \Pi}  \left[\beta^{(4)} +6\dot\beta \beta^{(3)}+4 \ddot\beta \left(\ddot\beta+3\dot\beta^2 \right) \right] -\frac{4}{\Phi \Pi^2} \left(\beta^{(3)} +6\dot\beta \ddot\beta \right)\left( 3\beta^{(3)} +10 \dot\beta \ddot\beta \right) \right\} \nonumber\\
&+6\left(2\ddot\beta +3\dot\beta^2  -\Lambda \right)=0.
\end{align}
It is noted that the existence of the third term in the left hand side of the  Euler-Lagrange equation \eqref{EL-2} is due to the fact that ${\cal L}$ contains not only $\beta$ and $\dot\beta$ but also $\ddot\beta$. 
Up to now, the desired  field equations \eqref{field-equation-1} and \eqref{field-equation-2} have been worked out thanks to the effective Euler-Lagrange equation approach. It is apparent that all of these field equations are higher order nonlinear ordinary differential equations. In particular, the Friedmann equation \eqref{field-equation-1}  is the third-order differential equation of $\beta$, while the other field equation \eqref{field-equation-2} is the fourth-order differential equation of $\beta$. 
Therefore, these field equations seems to be very difficult to be solved analytically. Fortunately, the studies done in Refs. \cite{barrow05,barrow06}, which are also about inflation in the fourth-order gravity, provide us a useful hint to figure out analytical solution to these field equations. As a result, detailed inflationary solutions will be presented in the next sections.
\section{Inflationary solutions}\label{sec3}
Following the Barrow-Hervik's papers in Refs. \cite{barrow05,barrow06} as well as our previous paper \cite{Do:2020vdc}, we will assume the following ansatz for the scale factor as 
\begin{equation} \label{ansatz}
\beta(t) =\zeta t,
\end{equation}
where $\zeta$ is a constant, whose value will be determined after solving the field equations. Consequently, the corresponding anticurvature scalar $A$ turns out to be
\begin{equation} \label{def-of-A-2}
A =\frac{4}{3 \zeta^2}.
\end{equation}
It is clear that $A$ is always singularity-free during an inflationary phase with $\zeta \gg1$. It is noted that the corresponding value of the Ricci scalar is given by
\begin{equation}
R = 12 \zeta^2.
\end{equation}
Hence, it is straightforward to have a relation
\begin{equation}
AR =16.
\end{equation}
As a result, the field equations \eqref{field-equation-1} and \eqref{field-equation-2} both reduce to the corresponding algebraic equation of $\zeta$,
\begin{equation} \label{equation-of-zeta}
27 \zeta^6 -9\Lambda \zeta^4  +9\alpha_1 \zeta^2 +16\alpha_2 =0.
\end{equation}
Setting $\hat \zeta \equiv \zeta^2 >0$ will lead the above equation to a cubic equation,
\begin{equation} \label{equation-of-hat-zeta}
27{\hat \zeta}^3 -9\Lambda \hat\zeta^2 +9\alpha_1 \hat \zeta +16 \alpha_2 =0.
\end{equation}
\subsection{Case 1: Vanishing $\alpha_2$}
It is noted that when setting $\alpha_2=0$, i.e., ignoring the contribution of the second order term $A^2$, we will arrive at an equation of $\hat \zeta$,
\begin{equation}
3{\hat \zeta}^2 -\Lambda \hat\zeta +\alpha_1=0,
\end{equation} 
which is nothing but that derived in our previous paper \cite{Do:2020vdc}. Furthermore, as pointed out in Ref. \cite{Do:2020vdc}, for the existence of an inflationary solution with $\zeta \gg1 $, or equivalently $\hat\zeta \gg 1$, $\alpha_1$ must be negative definite along with $|\alpha_1| \gg \Lambda \sim {\cal O}(1)$. Indeed, we can have an approximated solution for this equation such as
\begin{equation}
\hat\zeta \simeq \sqrt{\frac{-\alpha_1}{3}} \gg 1.
\end{equation}
\subsection{Case 2: Vanishing $\alpha_1$}
In this case, Eq. \eqref{equation-of-hat-zeta} simply becomes as
\begin{equation}
27{\hat \zeta}^3 -9\Lambda \hat\zeta^2 +16 \alpha_2 =0.
\end{equation}
It is clear to see that if $\alpha_2 <0$ then this equation always admits at least one positive root, $\hat \zeta >0$. On the other hand, it turns out that if $\alpha_2 <0$ this cubic equation will always admit only one real root along with the two other complex roots since its discriminant is always negative definite,
\begin{equation}
\Delta = -46656 \left[108 \alpha_2^2 -\Lambda^3 \alpha_2 \right] <0.
\end{equation}
In conclusion, in this case $\alpha_2$ should be negative, similar to $\alpha_1$ in the case 1, in order to have inflation. And the absolute value of $\alpha_2$ should be much larger than $\Lambda \sim {\cal O}(1)$ to have the following solution as
\begin{equation}
\hat\zeta \simeq \frac{2}{3} \left(-2\alpha_2 \right)^{1/3} \gg 1.
\end{equation}
\subsection{Case 3: Non-vanishing $\alpha_1$ and $\alpha_2$}
\begin{figure}[hbtp] 
\begin{center}
{\includegraphics[height=65mm]{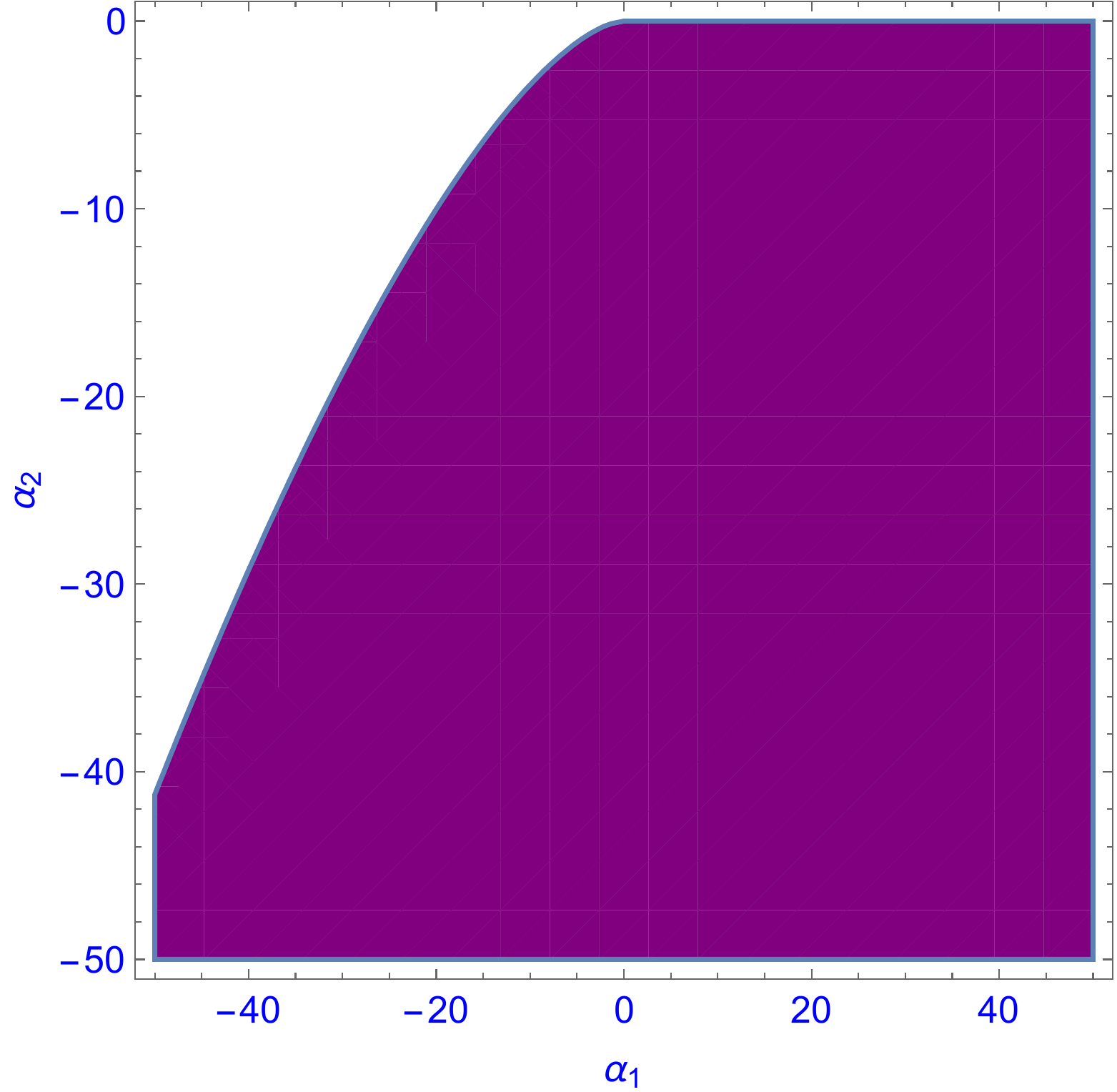}}\\
\caption{The purple region of both $\alpha_1$ and $\alpha_2$ for $\Lambda=1$, in which the cubic equation admits only one real, positive root $\hat\zeta$.}
\label{fig1}
\end{center}
\end{figure}

\begin{figure}[hbtp] 
\begin{center}
{\includegraphics[height=60mm]{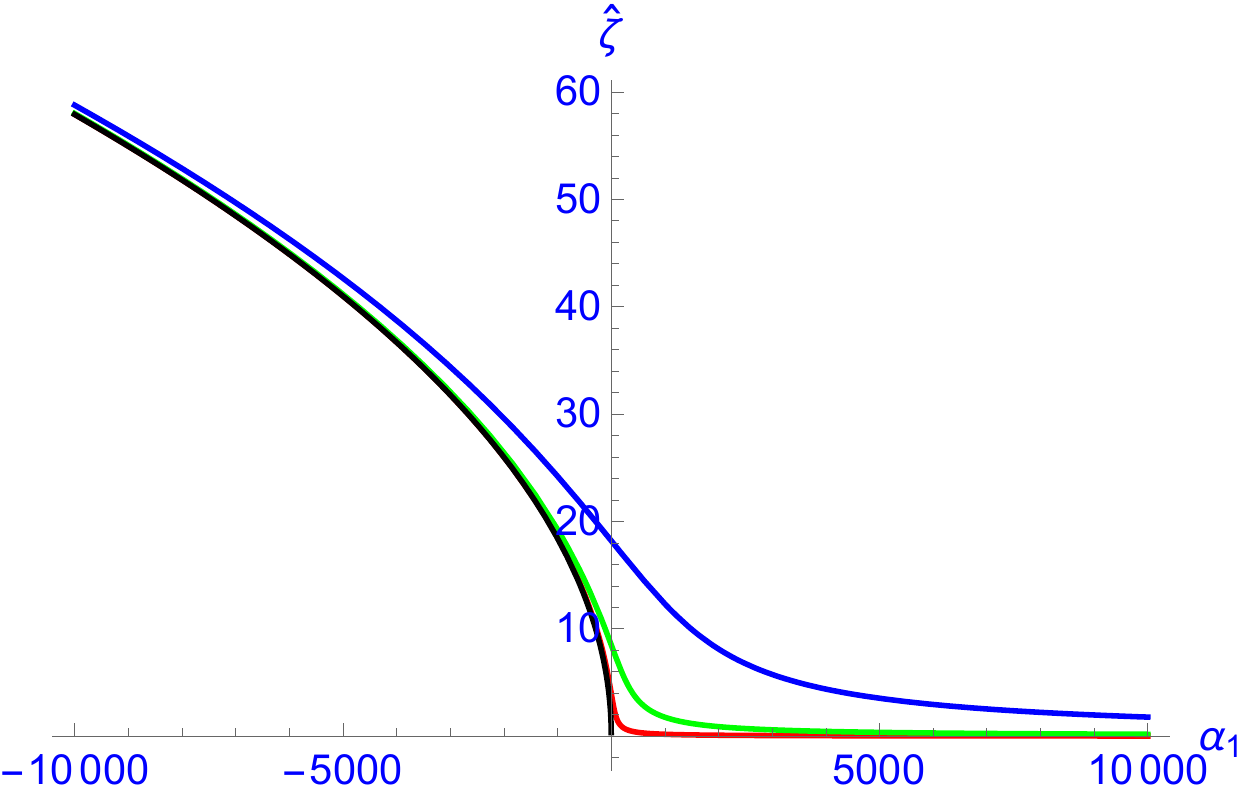}}\\
\caption{Behavior of the positive root $\hat\zeta$ for $\Lambda=1$ and several values of $\alpha_2$. In particular, the black, red, green, and blue curves correspond to $\alpha_2 =0$, $-100$, $-1000$, and -$10000$, respectively.}
\label{fig2}
\end{center}
\end{figure}

Now, we would like to consider a geneal scenario, in which $\alpha_1$ and $\alpha_2$ are both non-vanishing. In particular, we would like to see whether Eq. \eqref{equation-of-hat-zeta} with additional parameter $\alpha_2$ admits an inflationary solution to the present model. 
It is clear to see that if $\alpha_2 <0$ then this equation always admits at least one positive root, $\hat \zeta >0$, no matter the sign of $\alpha_1$. Furthermore, if one need only one real, positive root to this equation, which could represent an inflationary solution, an additional constraint should be mathematically satisfied, 
\begin{equation}
\Delta = -729 \left[108 \left(\alpha_1^3 +64\alpha_2^2 \right)+864 \Lambda \alpha_1 \alpha_2 -9\Lambda^2 \alpha_1^2 -64\Lambda^3 \alpha_2 \right] <0,
\end{equation}
where $\Delta$ is the following discriminant of this cubic equation. See Fig. \ref{fig1} for a constraint region of both $\alpha_1$ and $\alpha_2$ given that $\Lambda=1$, in which the cubic equation admits only one real, positive root $\hat\zeta$. According to this plot, we observe that there exists a wide region of positive $\alpha_1$, in which the corresponding $\hat\zeta$ will be positive definite, possible to be a desired inflationary solution.
As shown in Fig. \ref{fig2}, the contribution of the second order term $A^2$ will increase a little bit the value of scale factor $\hat\zeta$ when $\alpha_1 <0$ and $|\alpha_1| \gg 1$. However, there is an interesting point that it is possible to have an inflationary for positive $\alpha_1$. Furthermore, given a fixed negative value of $\alpha_2$, whose absolute value is assumed to be much larger than one, the smaller positive value of $\alpha_1$ is, the larger value $\zeta$ will be. For convenience, we classify all possibilities to have inflation into three sets of inequalities (I)-(III) for $\Lambda=1$: 
\begin{align}
\label{critical-1}
&{\rm (I)}: ~\alpha_1 <0, ~\alpha_2 <0,~ |\alpha_1| \gg |\alpha_2| \sim {\cal O}(1),\\
\label{critical-2}
&{\rm (II)}: ~\alpha_1<0, ~\alpha_2<0, ~|\alpha_{1}| \gg1 ~({\rm or} ~|\alpha_2| \gg 1),\\
\label{critical-3}
&{\rm (III)}: ~\alpha_1 >0,~ \alpha_2 <0,~ |\alpha_2| \gg \alpha_1 \sim {\cal O}(1).
\end{align}
In other words, one of these sets needs to be fulfilled if we would like to have inflation.
It seems that these inequalities are quite flexible to be fulfilled. Hence, we now face to a very important point that the inclusion of $A^2$ term would lead to the stability of the corresponding inflationary solutions or not. This issue will be investigated in detailed in the next section. 
\subsection{Minkowskian limit}
In this subsection, we would like to discuss a Minkowskian limit of this Ricci-inverse gravity model, which one might concern due to the existence of the anticurvature scalar $A$. In particular, one might worry that in a Minkowskian limit, $N\to 1$ and $\beta \to 0$, the anticurvature scalar $A$, whose general definition has been defined in Eq. \eqref{def-of-A-1}, would blow up and therefore the Ricci-inverse gravity model might lack the Minkowskian limit.  It should be noted that the possibility that the lack of Minkowskian limit might happen within the Ricci-inverse gravity model has already been mentioned, but without any detailed analysis, in the original paper \cite{Amendola:2020qho}. Fortunately, we will show that this is not the case for the solutions of the scale factor $\beta$ found in this paper as well as in our previous one \cite{Do:2020vdc}. Recall that for the ansatz of the scale factor chosen in Eq. \eqref{ansatz}, i.e., $\beta =\zeta t$, the corresponding value of the anticurvature scalar $A$ has been given by Eq. \eqref{def-of-A-2}, i.e., $A =4 \zeta^{-2} /3  = 4\hat \zeta ^{-1}/3$, provided that $N=1$. Additionally, the general equation of $\hat \zeta$ has been defined in Eq. \eqref{equation-of-hat-zeta}. In order to  discuss solely the Minkowskian limit of the Ricci-inverse gravity,  we will turn off the cosmological $\Lambda$. 

For the case 1 with the vanishing $\alpha_2$, it has been shown that $\hat\zeta = \sqrt{-\alpha_1/3}$, which leads to
\begin{equation}
\alpha_1 A = -4 \sqrt{\frac{-\alpha_1}{3}}. 
\end{equation}
Hence, it is obvious that the limit $\beta \to 0$ implies that $\hat\zeta \to 0$ and therefore $\alpha_1 \to 0$ and $\alpha_1 A \to 0$. In other words, $A$ will blow up but $\alpha_1 A$ will not as $\beta \to 0$. This result indicates that the solution found in this case always admits the Minkowskian limit. 

For the case 2 with the vanishing $\alpha_1$, it has appeared that $\hat\zeta = {2}\left(-2\alpha_2 \right)^{1/3}/3$. Consequently, we have
\begin{equation}
\alpha_2 A^2 = -2 \left(-2\alpha_2 \right)^{1/3}.
\end{equation}
Hence, it is clear that the limit $\beta \to 0$ leads to  $\hat\zeta \to 0$ and therefore $\alpha_2 \to 0$ and $\alpha_2 A^2 \to 0$. In other words, the solution found in the case 2 also admits the Minkowskian limit, similar to that found in the case 1.  

For the case 3, which is the general case with the non-vanishing $\alpha_1$ and $\alpha_2$, it turns out that 
\begin{equation} \label{min-1}
\alpha_1 A +\alpha_2 A^2 =  \frac{12\alpha_1 \hat\zeta +16\alpha_2}{9\hat\zeta^2}.
\end{equation}
On the other hand, we have, according to Eq.  \eqref{equation-of-hat-zeta}, that
\begin{equation} \label{min-2}
\alpha_1 A +\alpha_2 A^2  = - 3 \hat\zeta +\frac{\alpha_1}{3\hat\zeta}.
\end{equation}
Interestingly, this equation can be reduced to
\begin{equation}
\frac{3}{4}\alpha_1 A +\alpha_2 A^2 =-3\hat\zeta,
\end{equation}
which implies that both $\alpha_1 A $ and $\alpha_2 A^2$ will approach zero rather than blow up as $\hat\zeta \to 0$ if both $\alpha_1$ and $\alpha_2$ are assumed to be non-positive. In other words, it is possible to have the Minkowskian limit in the general case. 
\section{No-go theorem for inflationary solutions} \label{sec4}
In this section, we would like to investigate the stability of the isotropic inflationary solution within the extended Ricci-inverse gravity. As a result, we will finally reach to a no-go theorem for inflationary solutions of this proposed model due to the result that all these inflationary solutions will be shown to be unstable unexpectedly. 
\subsection{Dynamical system}
In cosmology, the stability analysis based on the dynamical system has been widely used \cite{barrow06,Bahamonde:2017ize}. Therefore, we will construct the corresponding dynamical system, which is nothing but a set of the first order differential equations called the autonomous equations, from the higher order field equations \cite{barrow06}. Note that this method has been used in our previous study on the original Ricci-inverse gravity model \cite{Do:2020vdc}. As a result, we have shown using this method that the original Ricci-inverse gravity admits unstable isotropic inflation.
First, we will introduce dynamical variables as follows \cite{barrow06,Do:2020vdc}
 \begin{align}
& B=\frac{1}{\dot\beta^2},\\
 &Q=\frac{\ddot\beta}{\dot\beta^2},\\
& Q_2 =\frac{\beta^{(3)}}{\dot\beta^3}, \\
& \Omega_\Lambda =\frac{\Lambda}{3\dot\beta^2},
  \end{align}
 here the Hubble constant is given by $H=\dot\beta$. Note that we no longer have the  shear terms like $\Sigma =\dot\sigma/\dot\beta$ since the parameter $\sigma(t)$ representing spatial anisotropies vanishes for isotropic metrics.
 As a result, the corresponding set of autonomous equations of dynamical variables turn out to be
 \begin{align}
  \label{Dyn-1}
 B' &= -2QB,\\
  \label{Dyn-2}
 \Omega_\Lambda' &= -2Q\Omega_\Lambda,\\
  \label{Dyn-3}
 Q' &=Q_2 -2Q^2,\\
 \label{Dyn-4}
 Q_2 '&= \frac{\beta^{(4)}}{\dot\beta^4}-3Q Q_2,
 \end{align}
 where $' \equiv d/d\tau$ with $\tau =\int \dot\beta dt$ is nothing but the dynamical time variable. It is noted that the term ${\beta^{(4)}}/{\dot\beta^4}$ in Eq. \eqref{Dyn-4} can be figured out from the field equation \eqref{field-equation-2}, which can be written in terms of the dynamical variables as follows
 \begin{align} \label{Dyn-5}
&\frac{\alpha_1}{3\Phi} \left\{ 3B -\frac{1}{\Phi} \left(Q +3 \right) +\frac{2}{B\Phi^2}  \left[\frac{\beta^{(4)}}{\dot\beta^4}+6Q_2 +2 Q \left(Q+4 \right) \right] - \frac{6}{B^2\Phi^3} \left(Q_2 +2Q \right)^2 \right\} \nonumber\\
&+\frac{3\alpha_1}{\Pi} \left[6B+ \frac{2}{B\Pi^2} \left(\frac{\beta^{(4)}}{\dot\beta^4}+6Q_2+6 Q^2 \right) - \frac{6}{B^2\Pi^3} \left(Q_2+6Q\right)^2 \right] \nonumber\\
&+\frac{\alpha_2 }{9\Phi^2}\left\{ 3B -\frac{2}{\Phi} \left( Q+ 3 \right) +\frac{6}{B\Phi^2} \left[ \frac{\beta^{(4)}}{\dot\beta^4} +6Q_2 + 2 Q \left( Q+ 4  \right)\right] -\frac{24}{B^2\Phi^3} \left(Q_2 +2 Q \right)^2  \right\} \nonumber\\
&+\frac{27\alpha_2}{\Pi^2} \left[ 3B +\frac{2}{B\Pi^2} \left(\frac{\beta^{(4)}}{\dot\beta^4}+6 Q_2 +6Q^2   \right)  -\frac{8 }{B^2 \Pi^3}  \left(Q_2 +6Q \right)^2 \right] \nonumber\\
& +\frac{\alpha_2}{\Pi \Phi} \left\{ 3B -\frac{2}{\Phi} \left(Q +3 \right) +\frac{4}{B\Phi^2} \left[\frac{\beta^{(4)}}{\dot\beta^4} +6Q_2+2 Q \left(Q +4 \right) \right] -\frac{12}{B^2\Phi^3} \left(Q_2 +2 Q \right)^2 \right. \nonumber\\
&\left.+\frac{2}{B\Pi \Phi}  \left[\frac{\beta^{(4)}}{\dot\beta^4} +6Q_2 +4 Q\left(Q+3 \right) \right] -\frac{4}{B^2\Pi \Phi^2} \left(Q_2 +2 Q \right)\left( 3 Q_2  +14 Q \right) \right\} \nonumber\\
&+\frac{\alpha_2}{\Phi \Pi} \left\{ 9B +\frac{4}{B\Pi^2}  \left(\frac{\beta^{(4)}}{\dot\beta^4} +6Q_2 +6Q^2 \right) -\frac{12}{B^2\Pi^3} \left(Q_2+6Q \right)^2 \right. \nonumber\\
&\left. +\frac{2}{B\Phi \Pi}  \left[ \frac{\beta^{(4)}}{\dot\beta^4} +6Q_2+4 Q \left(Q+3 \right) \right] -\frac{4}{B^2\Phi \Pi^2} \left(Q_2 +6Q \right)\left( 3Q_2 +10 Q \right) \right\} \nonumber\\
&+6\left(2Q +3 -3\Omega_\Lambda \right)=0.
\end{align}
where $\Phi$ and $\Pi$ are now functions of the dynamical variables such as
\begin{align}
\Phi =\frac{1}{{B}} \left(Q+1 \right),~\Pi =\frac{1}{B} \left(Q+3\right).
\end{align}
Before going to solve fixed point to the dynamical system, we would like to note that there is a constraint equation, which is nothing but the Friedmann equation \eqref{field-equation-1} written in terms of the dynamical variables as
\begin{align}  \label{Dyn-6}
&\frac{\alpha_1}{3\Phi} \left[ 3B -\frac{1}{\Phi} \left(Q +3 \right) +\frac{2}{B\Phi^2}  \left(Q_2 +2Q \right) \right] +\frac{3\alpha_1}{\Pi} \left[2B + \frac{2}{B\Pi^2}\left(Q_2+6Q \right) \right] \nonumber\\
&+\frac{\alpha_2 }{9\Phi^2}\left[ 5B -\frac{2}{\Phi} \left(Q+3 \right) +\frac{6}{B\Phi^2} \left(Q_2  + 2 Q \right)  \right] +\frac{9\alpha_2}{\Pi^2} \left[ 3B +\frac{6}{B\Pi^2} \left(Q_2 +6Q \right) \right] \nonumber\\
&+\frac{\alpha_2}{\Pi \Phi} \left[5B -\frac{2}{\Phi} \left(Q +3 \right) +\frac{4}{B\Phi^2}  \left(Q_2 +2Q \right) +\frac{2}{B\Pi \Phi} \left(Q_2+4 Q \right) \right] \nonumber\\
& +\frac{\alpha_2}{\Phi \Pi} \left[3B +\frac{4}{B\Pi^2}  \left(Q_2 +6Q \right) +\frac{2}{B\Phi \Pi}  \left(Q_2 +4Q \right) \right] \nonumber\\
&+ 2\left(3  -3\Omega_\Lambda\right)=0.
\end{align}
Hence, all found fixed point solutions should satisfy this important constraint equation.
\subsection{Fixed point}
Following the previous works \cite{barrow06,Do:2020vdc}, we are going to figure out the corresponding fixed point of the dynamical system described by the autonomous equations \eqref{Dyn-1}, \eqref{Dyn-2}, \eqref{Dyn-3}, \eqref{Dyn-4}, \eqref{Dyn-5}, and \eqref{Dyn-6}. Mathematically, the fixed point is a solution of the following equations
\begin{equation}
B'=\Omega_\Lambda' =Q'=Q_2'=0.
\end{equation}
As a result, these equations lead to
\begin{equation}
Q_2=Q =0
\end{equation}
along with an equation of $B^2$,
\begin{equation}\label{equation-of-B} 
16 \alpha_2 B^3 +9\alpha_1 B^2 -27 \left(\Omega_\Lambda-1\right)=0.
\end{equation}
More interestingly, this cubic equation can be reduced to
\begin{equation} 
27\dot\beta^6 -9\Lambda \dot\beta^4 +9\alpha_1 \dot\beta^2 +16\alpha_2 =0,
\end{equation}
which is nothing but Eq. \eqref{equation-of-zeta} given that $\dot\beta =\zeta$. This result implies that the exponential solution found in the previous section is equivalent to the isotropic fixed point found in this section. Therefore, investigating the stability of the fixed point will yield the stability of the exponential solution.

It is apparent that   $B=\dot\beta^{-2}$ and $\Omega_\Lambda = \Lambda \dot\beta^{-2}/3$ both must be much smaller than one for an inflationary solution due to the following constraint that $\dot\beta \gg 1$. Consequently, Eq. \eqref{equation-of-B} indicates that
\begin{equation} \label{inequality-for-inflation}
16 \alpha_2 B^3 +9\alpha_1 B^2 +27 = 27 \Omega_\Lambda \ll 1,
\end{equation}
or equivalently,
\begin{equation}
16 \alpha_2 B^3 +9\alpha_1 B^2 \simeq - 27 <0.
\end{equation}
This result implies that at least either $\alpha_1$ or $\alpha_2$  is negative definite and has absolute value much larger than one. This requirement can be fulfilled by one of three sets of inequalities (I)-(III) shown in Eqs. \eqref{critical-1}, \eqref{critical-2}, and \eqref{critical-3}, respectively. To be more specific, we will plot in the Fig. \ref{fig3} the green region of $\alpha_1 B^2 $ and $\alpha_2 B^3$ with $27\Omega_\Lambda$ is chosen to be $10^{-3}$, where any real solution of Eq. \eqref{equation-of-B} should belong to. 
\subsection{Stability analysis of  fixed point}
Similar to the previous investigations \cite{barrow06,Do:2020vdc}, we are going to perturb the autonomous equations around the fixed point to see whether or not the unstable mode(s) exists. As a result, a set of perturbed equations is given by
\begin{align}
\label{iso-pert-1}
\delta B' &=-2B \delta Q,\\
\label{iso-pert-2}
\delta \Omega_\Lambda'&=-2\Omega_\Lambda \delta Q,\\
\label{iso-pert-3}
\delta Q' &= \delta Q_2,\\
\label{iso-pert-4}
\delta Q_2' &= \delta \left(\frac{\beta^{(4)}}{\dot\beta^4}\right),
\end{align}
where $\delta \left({\beta^{(4)}}/{\dot\beta^4}\right)$  will be figured out from the following perturbed equation,
 \begin{align} \label{perturb-1}
&\frac{\alpha_1}{3\Phi} \left\{ 3 \delta B + \frac{3}{\Phi^2}\delta \Phi -\frac{1}{\Phi} \delta Q+\frac{2}{B\Phi^2}  \left[ \delta \left(\frac{\beta^{(4)}}{\dot\beta^4} \right)+6 \delta Q_2 +8\delta Q  \right]  \right\} +\frac{3\alpha_1}{\Pi} \left\{ 6 \delta B -\frac{18}{\Pi^2} \delta \Pi + \frac{2}{B\Pi^2} \left[\delta \left(\frac{\beta^{(4)}}{\dot\beta^4}\right)+6\delta Q_2 \right] \right\} \nonumber\\
&+\frac{\alpha_2 }{9\Phi^2}\left\{ 3\delta B + \frac{12}{\Phi^2} \delta \Phi -\frac{2}{\Phi}\delta Q  +\frac{6}{B\Phi^2} \left[ \delta \left(\frac{\beta^{(4)}}{\dot\beta^4}\right) +6\delta Q_2 + 8\delta Q \right]  \right\} +\frac{27\alpha_2}{\Pi^2} \left\{ 3\delta B -\frac{18}{\Pi^2}\delta \Pi+\frac{2}{B\Pi^2} \left[ \delta \left(\frac{\beta^{(4)}}{\dot\beta^4}\right)+6\delta Q_2   \right]   \right\} \nonumber\\
& +\frac{\alpha_2}{\Pi \Phi} \left\{ 3\delta B  +\frac{9}{\Phi^2} \delta \Phi +\frac{3}{\Pi \Phi} \delta \Pi -\frac{2}{\Phi}\delta Q +\frac{4}{B\Phi^2} \left[ \delta \left(\frac{\beta^{(4)}}{\dot\beta^4}\right) +6\delta Q_2+8 \delta Q  \right]  +\frac{2}{B\Pi \Phi}  \left[ \delta \left(\frac{\beta^{(4)}}{\dot\beta^4} \right)+6 \delta Q_2 +12\delta Q \right]  \right\} \nonumber\\
&+\frac{\alpha_2}{\Phi \Pi} \left\{ 9\delta B -\frac{27}{\Phi \Pi} \delta \Phi -\frac{27}{ \Pi^2}\delta \Pi +\frac{4}{B\Pi^2}  \left[ \delta \left(\frac{\beta^{(4)}}{\dot\beta^4} \right) +6\delta Q_2  \right]  +\frac{2}{B\Phi \Pi}  \left[\delta \left( \frac{\beta^{(4)}}{\dot\beta^4} \right) +6 \delta Q_2+12 \delta Q \right]  \right\} \nonumber\\
&+6\left(2\delta Q -3 \delta \Omega_\Lambda \right)=0,
\end{align}
where
\begin{align} \label{rel-1-1}
&\delta \Phi = -\frac{1}{B^2}\delta B +\frac{1}{B}\delta Q,\\
 \label{rel-1-2}
&\delta \Pi = -\frac{3}{B^2}\delta B +\frac{1}{B}\delta Q,
\end{align}
along with
\begin{equation}\label{rel-2}
\Phi = \frac{1}{B}, ~\Pi = \frac{3}{B}.
\end{equation}
This perturbed equation is derived from the field equation \eqref{Dyn-5}.
It is noted that the perturbed Friedmann equation turns out to be
\begin{align}  \label{perturb-2}
&\frac{\alpha_1}{3\Phi} \left[ 3 \delta B +\frac{3}{\Phi^2}\delta \Phi -\frac{1}{\Phi}\delta Q  +\frac{2}{B\Phi^2}  \left(\delta Q_2 +2\delta Q \right) \right] +\frac{3\alpha_1}{\Pi} \left[2\delta B -\frac{6}{\Pi^2}\delta \Pi + \frac{2}{B\Pi^2}\left(\delta Q_2+6 \delta Q \right) \right] \nonumber\\
&+\frac{\alpha_2 }{9\Phi^2}\left[ 5\delta B + \frac{8}{\Phi^2} \delta \Phi - \frac{2}{\Phi}\delta Q  +\frac{6}{B\Phi^2} \left(\delta Q_2  + 2\delta Q \right)  \right] +\frac{9\alpha_2}{\Pi^2} \left[ 3 \delta B -\frac{18}{\Pi^2} \delta \Pi+\frac{6}{B\Pi^2} \left(\delta Q_2 +6 \delta Q \right) \right] \nonumber\\
&+\frac{\alpha_2}{\Pi \Phi} \left[5 \delta B + \frac{7}{\Phi^2} \delta \Phi +\frac{1}{\Pi \Phi}\delta \Pi -\frac{2}{\Phi}\delta Q +\frac{4}{B\Phi^2}  \left(\delta Q_2 +2\delta Q \right) +\frac{2}{B\Pi \Phi} \left(\delta Q_2+4\delta Q \right) \right] \nonumber\\
& +\frac{\alpha_2}{\Phi \Pi} \left[3\delta B -\frac{9}{\Phi \Pi} \delta \Phi -\frac{9}{\Pi^2}\delta \Pi +\frac{4}{B\Pi^2}  \left(\delta Q_2 +6\delta Q \right) +\frac{2}{B\Phi \Pi}  \left(\delta Q_2 +4\delta Q \right) \right] -6 \delta \Omega_\Lambda=0.
\end{align}
Thanks to the useful relations shown in Eqs. \eqref{rel-1-1}, \eqref{rel-1-2}, and \eqref{rel-2}, Eq. \eqref{perturb-2} can be solved to give
\begin{equation}
\delta \Omega_\Lambda = \frac{2B}{81} \left[ 9 \left(8 \alpha_2 B+ 3\alpha_1   \right)\delta B  + 2B \left(11\alpha_2 B+3\alpha_1   \right) \left(3\delta Q+ \delta Q_2 \right)\right].
\end{equation}
By inserting this solution into Eq. \eqref{perturb-1} we are able to obtain the following value of $\delta \left({\beta^{(4)}}/{\dot\beta^4}\right)$ as 
\begin{equation}
\delta \left(\frac{\beta^{(4)}}{\dot\beta^4}\right) =\frac{3\left(32 \alpha_2 B^3+9\alpha_1 B^2 -27  \right)}{2B^2 \left(11\alpha_2 B+3\alpha_1 \right)}\delta Q  - 3 \delta Q_2 .
\end{equation}
By taking exponential perturbations,
\begin{align}
\delta B&=A_1 \exp\left[\mu \tau \right],\\
\delta \Omega_\Lambda &=A_2 \exp\left[\mu \tau \right], \\
\delta Q&=A_3 \exp\left[\mu \tau \right], \\
\delta Q_2& =A_{4}\exp\left[\mu \tau \right],
\end{align}
it is possible to write all perturbation equations \eqref{iso-pert-1}, \eqref{iso-pert-2}, \eqref{iso-pert-3}, and \eqref{iso-pert-4} as a matrix equation,
 \begin{equation} \label{stability-equation}
{\cal M}\left( {\begin{array}{*{20}c}
   A_1  \\
   A_{2}  \\
   A_3  \\
   A_{4}\\
 \end{array} } \right) \equiv \left[ {\begin{array}{*{20}c}
   {\mu} & {0} & {2B } & {0 }   \\
   {0 } & {\mu} & {2\Omega_\Lambda} & {0}  \\
   {0} & {0} & {\mu } & {-1} \\
   {0}&{0}&{-\frac{3\left(32 \alpha_2 B^3+9\alpha_1 B^2 -27  \right)}{2B^2 \left(11\alpha_2 B+3\alpha_1 \right)}} &{\mu+3} \\
 \end{array} } \right]\left( {\begin{array}{*{20}c}
   A_1  \\
   A_{2}  \\
   A_3  \\
   A_{4}\\
 \end{array} } \right) = 0.
\end{equation}
Mathematically, this matrix equation admits non-trivial solutions if and only if
\begin{equation}
\det {\cal M}=0,
\end{equation}
which can be determined to be
\begin{equation}
\mu^2 \left[2B^2 \left(11\alpha_2 B +3\alpha_1 \right)\mu^2 +6B^2 \left(11\alpha_2 B +3\alpha_1 \right)\mu -96\alpha_2 B^3 -27\alpha_1 B^2 +81 \right]=0.
\end{equation}
As a result, besides two trivial eigenvalues, $\mu_{1,2} =0$, this equation admits two non-trivial ones given by
\begin{equation} \label{mu3,4}
\mu_{3,4}=- \frac{1}{2} \left[3 \mp \sqrt{\frac{3 \left(97\alpha_2 B^3 +27\alpha_1 B^2-54\right)}{B^2 \left(11\alpha_2 B +3\alpha_1 \right)}}\right].
\end{equation}
Note that a stable inflationary solution happens only when all obtained $\mu_{3,4}$ or their real part (if they are complex number) turn out to be non-positive definite. Otherwise, unstable mode(s) to the inflationary solution  will arise accordingly. 
It is straightforward to see that if we set $\alpha_2=0$, or equivalently neglecting the contribution of $A^2$, the eigenvalues $\mu_{3,4}$ all reduce to
\begin{equation}
\mu_{3,4} \to \bar\mu_{3,4}=- \frac{3}{2} \left[1 \mp \sqrt{3 - \frac{ 6}{\alpha_1 B^2 }}\right],
\end{equation}
which are nothing but $\mu_{6,7}$ found in our previous works \cite{Do:2020vdc} provided a replacement that $\alpha_1 \to \alpha$. It is clear in this case that $\bar\mu_3$ is always positive for $\alpha_1<0$, making the inflationary solution unstable  \cite{Do:2020vdc}.

One can ask what will happen if $\alpha_1=0$ and $\alpha_2 \neq 0$. It turns out that the eigenvalues $\mu_{3,4}$ now become as
\begin{equation}
\mu_{3,4} \to  \hat\mu_{3,4}=- \frac{3}{2} \left[1 \mp \sqrt{\frac{97}{33}-\frac{18}{11\alpha_2 B^3}}\right].
\end{equation}
It is clear that  $\hat\mu_3$ is always positive for $\alpha_2 <0$  then  the corresponding inflationary solution is unstable too.

\begin{figure}[hbtp] 
\begin{center}
{\includegraphics[height=60mm]{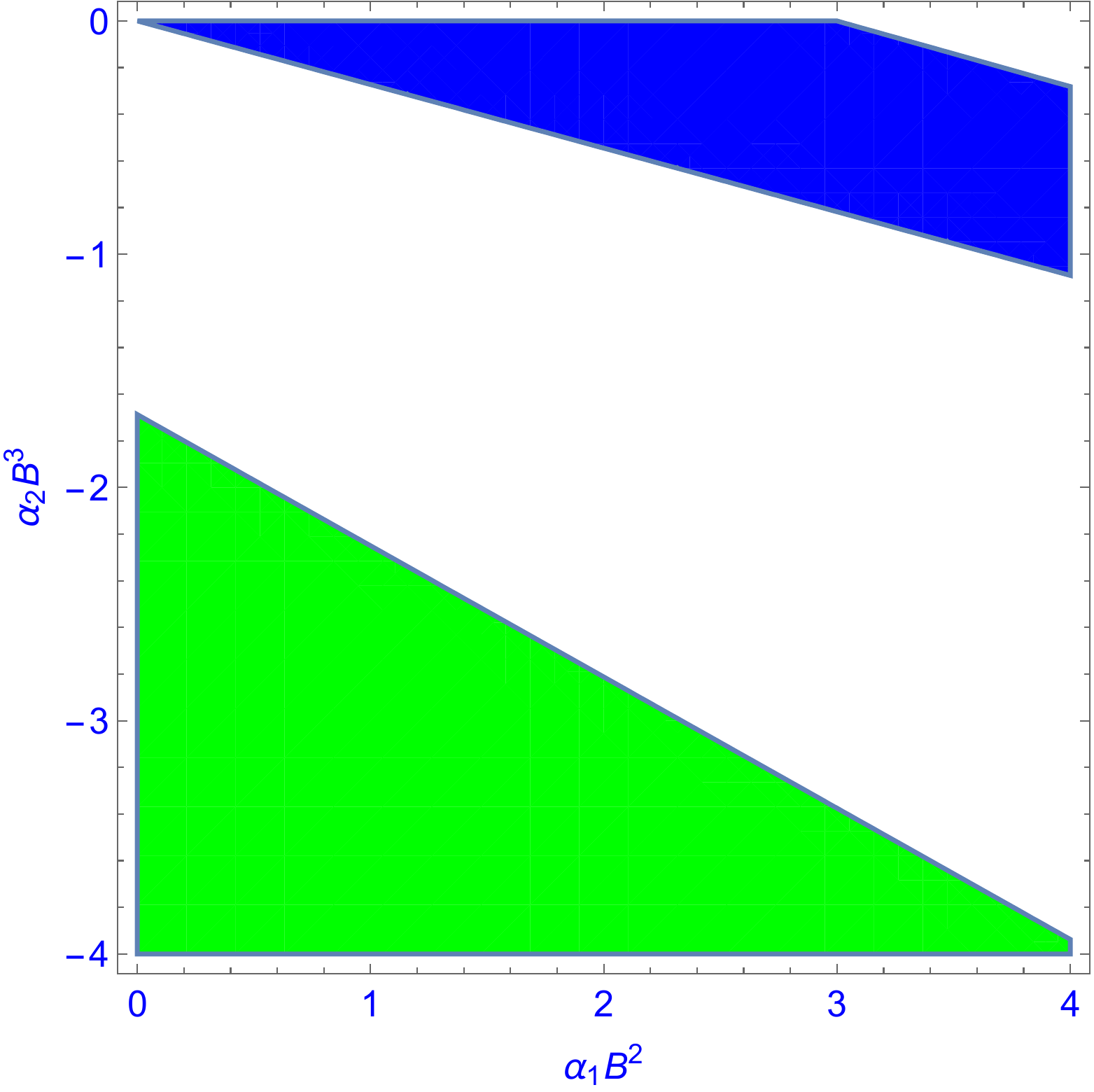}}\\
\caption{The stability region is colored as blue, while the existence region of real solution of Eq. \eqref{equation-of-B} is colored as green. It is clear that these two colored regions do not overlap each other, meaning no stable isotropic inflation exists in the extended model.}
\label{fig3}
\end{center}
\end{figure}

Now, we would like to examine the contribution of $\alpha_2 A^2$ in terms of $\alpha_2 B^3$ to see whether the eigenvalues $\mu_{3,4}$ defined in Eq. \eqref{mu3,4} with both non-vanishing $\alpha_1$ and $\alpha_2$ act as stable modes. We expect that the appearance of $\alpha_2$ in Eq. \eqref{mu3,4} would leave extra space for the existence of stable modes of the inflationary solution. As a result, the non-positivity of $\mu_{3,4}$ addresses the following inequality
\begin{equation} \label{inequality-for-stability}
\frac{ 97\alpha_2 B^3 +27\alpha_1 B^2-54}{B^2 \left(11\alpha_2 B +3\alpha_1 \right)}\leq 3.
\end{equation}
It turns out that for a stable inflationary solution it should satisfy not only the the inequality \eqref{inequality-for-stability} but also one of three sets of inequalities  (I), (II), and (III) described by Eqs. \eqref{critical-1}, \eqref{critical-2}, and \eqref{critical-3}, respectively. Therefore, we will numerically examine whether the inequality \eqref{inequality-for-stability} is satisfied or not in one of three different regions (I), (II), and (III). It turns out that only the region (III) is suitable for the existence of stable modes. To be more specific, we plot in Fig. \ref{fig3} the blue region for the inequality \eqref{inequality-for-stability}, where any real fixed point solution will be stable. However, one can wonder that is there any inflationary solution existing in this blue region. To address this question, we also plot in Fig. \ref{fig3} the green region for the existence of inflationary solution, following the inequality shown in Eq. \eqref{inequality-for-inflation} with $27\Omega_\Lambda$ is chosen to be $10^{-3}$. It appears that these two colored regions do not have common points, meaning that the inflationary solution found in this extended Ricci-inverse gravity model will no longer be stable as expected. This result unexpectedly raises more doubt about the cosmological validity of the Ricci-inverse gravity. 
\section{Conclusions} \label{final} 
We have proposed an extension of the Ricci-inverse gravity model \cite{Amendola:2020qho} by introducing the second order term $A^2$ as a correction. As a result, we have been able to derive the homogeneous and isotropic inflation within this extension. Unfortunately, the no-go theorem based on the stability analysis has been achieved for this  isotropic inflation. As a result, this theorem implies that it is impossible to have a stable isotropic inflation in this extended Ricci-inverse gravity model.  This result together with our previous investigation \cite{Do:2020vdc} raise more doubt about the implication of the Ricci-inverse gravity for inflationary phase of our universe. This result also indicates that the isotropic inflation might be not suitable for inflationary phase of our universe within the context of the Ricci-inverse gravity. An investigation on anisotropic inflation within this extended Ricci-inverse gravity model is therefore necessary \cite{Do:2020vdc}. Of course, other possible types of $f(A)$ or $f(R,A)$ of the Ricci-inverse gravity like that proposed in Ref. \cite{Amendola:2020qho} should also be examined.  We will leave these issues  for our further studies. We hope that our present study would be useful for other studies of the cosmological implications of the Ricci-inverse gravity.
\begin{acknowledgments}
The author would like to thank the referee very much for useful comments.
The author would also like to thank Dr. Sunny Vagnozzi and Dr. Alberto Salvio very much for introducing some other interesting features of the fourth-order gravity. This study is supported by the Vietnam National Foundation for Science and Technology Development (NAFOSTED) under grant number 103.01-2020.15. 
\end{acknowledgments}

\end{document}